# The Phenomenology of Aftershocks

A.V. Guglielmi[1], B.I. Klain[2]

[1] *Schmidt Institute of Physics of the Earth, Russian Academy of Sciences, Moscow, Russia, guglielmi@mail.ru*
[2] *Borok Geophysical Observatory, the Branch of Schmidt Institute of Physics of the Earth, Russian Academy of Sciences, Borok, Yaroslavl Region, Russia, klb314@mail.ru*

**Abstract**

The presented paper is devoted to the search for mathematical basis for describing the aftershock evolution of strong earthquakes. We consider the experimental facts and heuristic arguments that allow to make a choice and to focus on the nonlinear diffusion equation as the master equation. Analysis of the master equation indicates that, apparently, the selected mathematical basis makes it possible to simulate two important properties of the aftershock evolution known from the experiment. We are talking about the Omori law and the slow propagation of aftershocks from the epicenter of the main shock.

**Contents**

1. Introduction

Quite recently, at the international conference "Problems of Geocosmos–2018" in St. Petersburg, a report [1] was made, which did not attract much attention of the conference participants. Meanwhile, the result of the experimental study presented in the report is interesting and moreover rather mysterious. The authors of [1] found that the epicenters of aftershocks spread from the epicenter of the main shock at a velocity of approximately several kilometers per hour. This velocity is in some sense an intermediate. Indeed, the propagation rate of aftershocks is several orders higher than the rate of slow deformation waves known in geotectonics, and several orders lower than the velocities of seismic waves. Thus, we are faced with the problem of interpretation of the phenomenon discovered in [1] (see details in [2]).

The problem of physical understanding and mathematical description of aftershocks has a rich history (e.g., see [3, 4]). Even in the late 19th century Fusakichi Omori [5] proposed to describe the evolution of aftershocks phenomenologically using the formula

$$n(t) = \frac{k}{c+t}. \tag{1}$$

Here $n$ is the frequency of aftershocks in the epicentral zone of a strong earthquake, and $t$ is the time, $t \geq 0$. The phenomenological parameters $k > 0$ and $c > 0$ are determined from seismic data. Formula (1) is known in the physics of earthquakes as the Omori law [3–6]. Hirano [7], Utsu and others [8] proposed to improve formula (1) by introducing one more (third) parameter into it. Another approach to the problem is to replace formula (1) with a differential equation, which contains only one phenomenological parameter:

$$\frac{dn}{dt} + \sigma n^2 = 0. \tag{2}$$

Here $\sigma$ is the deactivation coefficient of the earthquake source which "cooling down" after the main shock [9].

It would seem that replacing (1) with (2) gives nothing new. Indeed, the general solution of equation (2) coincides with formula (1) up to notation. However, the methodological advantage of the evolution equation (2) in comparison with formula (1) becomes obvious as soon as we try to go beyond the framework of the model that was implicitly at the heart of formula (1). After all, formula (1) is valid under the mandatory condition $\sigma = \text{const}$. In other words, it was assumed that the state of the earthquake source characterized by the parameter $\sigma$ remains stationary during the evolution of aftershocks. Meanwhile, from heuristic considerations, it seems obvious that in reality the deactivation coefficient can change over time, since after the formation of a main rupture, the geological environment relaxes to a new state of equilibrium. The rocks in the earthquake source are clearly in a non-equilibrium state after the main shock. Thus, a natural generalization of Omori's law (1) is the formula

$$n(t) = n_0 \left[ 1 + n_0 \int_0^t \sigma(t') dt' \right]^{-1}. \quad (3)$$

For $\sigma = \text{const}$, formula (3) coincides with the Omori formula (1), $k = 1/\sigma$, and $c = 1/n_0\sigma$.

We will consider other interesting generalizations of Eq. (2) in Section 2. In Section 3, we will find the conditions for the Omori law to apply. Section 4 is devoted to the propagation of aftershocks found in [1]. In Section 5, we will discuss the issues of a phenomenological description of aftershocks and touch upon the problem of describing foreshocks and main shock. In the final section 6, we will summarize some preliminary results.

## 2. Master equation

We will try to derive, or rather guess, the differential equation the solutions of which, on the one hand, agree with the Omori law, and on the other hand, imitate the process of horizontal propagation of aftershocks found in [1].

As a first step, we will take into account the spatial heterogeneity of the distribution of aftershocks in the epicentral zone, i.e., replace $n(t)$ with $n(\mathbf{x},t)$, where $\mathbf{x}$ is a 2D radius vector that sets the position of a point on the earth's surface relative to the epicenter of the main shock. The next step is the most critical, since we need to choose the type of partial differential equation, presumably describing the space-time evolution of aftershocks. A careful analysis of the observational data led us to the idea that we, most likely, should be dealing not with a hyperbolic, but with a parabolic equation. Additional information contained in the equation of evolution of aftershocks (2), averaged over the epicentral zone, suggests that it is reasonable to choose as a sample the equation of nonlinear diffusion [10], known in the mathematical and natural science literature as the Kolmogorov-Petrovsky-Piskunov equation, or abbreviated KPP-equation:

$$\partial n / \partial t = D \nabla^2 n + F(n). \quad (4)$$

Here $D$ is the diffusion coefficient, $\nabla$ is the 2D Hamilton operator. The KPP-equation is widely used in biology and chemistry [11–14], in astrophysics [15], as well as in geotectonics for description of slow deformation waves in the lithosphere of the Earth (see [16] and the literature cited there).

We expand the function $F(n)$ in a power series and restrict ourselves to the first two terms: $F(n) = \gamma n - \sigma n^2$. The first term describes in its simplest form the fact that after the main shock the earthquake source is not only a nonequilibrium, but also an unstable dynamical system. We have kept the second term, focusing on the Omori law in the form (2). The result is a master equation

$$\partial n / \partial t = n(\gamma - \sigma n) + D\nabla^2 n, \qquad (5)$$

presumably describing the process of propagation of aftershocks.

Further generalization consists in replacing the scalar $D$ with a 2D tensor

$$\hat{D} = \begin{Vmatrix} D_\parallel & 0 \\ 0 & D_\perp \end{Vmatrix} \qquad (6)$$

in order to take into account the anisotropy of the fault system in the epicentral zone. However, we will not dwell on this.

In conclusion of this section we recall the derivation of the KPP equation (see for example [13]). We restrict ourselves to the one-dimensional case and write down a rather general integro-differential equation

$$\frac{\partial n}{\partial t} = \Phi(n) + \int_{-\infty}^{\infty} K(x-y) \cdot n(y,t) \cdot dy \qquad (7)$$

with a symmetric kernel $K(x-y) = K(y-x)$. If $K \to 0$ for $|x-y| \to \infty$, then expanding $n(x-z,t)$ into a Taylor series in powers of $x$, we obtain

$$\frac{\partial n}{\partial t} = \gamma n + \Phi(n) + D\frac{\partial^2 n}{\partial x^2} + \ldots , \qquad (8)$$

where

$$\gamma = \int_{-\infty}^{\infty} K(z) \cdot dz, \quad D = (1/2) \cdot \int_{-\infty}^{\infty} z^2 K(z) \cdot dz, \qquad (9)$$

and $z = x - y$. Let us $\Phi(n) = -\sigma n^2$, and confine the first two terms in the series. In this approximation we obtain equation (5) from which after phenomenological reduction follows the Omori law in the form (2).

3. **Omori law**

Equation (5) contains three phenomenological parameters, $\gamma$, $\sigma$ and $D$. Let us assume that the parameter $D$ is so small that the diffusion term in the equation can be neglected:

$$dn / dt = n(\gamma - \sigma n). \qquad (10)$$

We got the logistic equation (Verhulst equation [17]). It can be used to evaluate the conditions of applicability of the Omori law (see below). One can also make further generalization by adding a random function simulating seismic noise to the right-hand side of (10). In this case, the dynamic equation (10) will become a stochastic equation.

A more rigorous approach consists in averaging (5) over the epicentral zone. We denote the averaging procedure by angle brackets: $<n(\mathbf{x},t)>=n(t)$. There is a fairly wide class of functions for which $<\nabla^2 n(\mathbf{x},t)>=0$. However, when studying a nonlinear diffusion process, the transition from (5) to (10) by averaging requires a more subtle analysis in view of the fact that in general $<n^2>\neq<n>^2$.

If the parameters $\gamma$ and $\sigma$ are positive constants, then equation (10) can be easily integrated. Let us introduce the notations $n_\infty = \gamma/\sigma$, and

$$t_\infty = \frac{1}{\gamma}\ln\left(1-\frac{n_\infty}{n_0}\right), \qquad (11)$$

where $n_0 = n(0)$. Then, for $n \geq 0$ the solutions of the Verhulst equation will have the form shown in Figure 1. The monotonically increasing lower branch of the logistic function is often used in biology, but it is not suitable for describing the evolution of aftershocks for the obvious reason. So let's focus on the top branch.

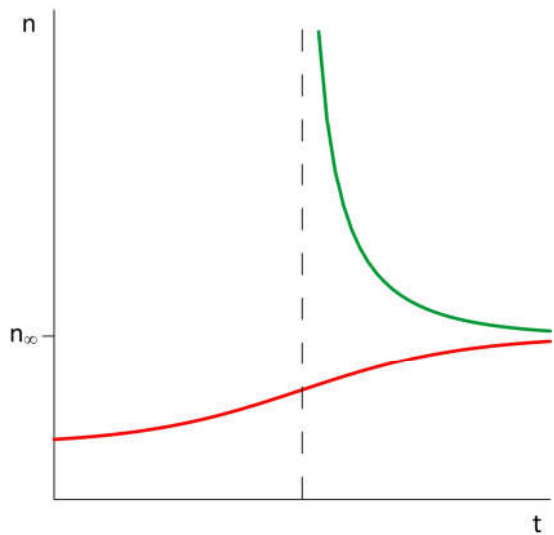

Fig. 1. Graphic representation of solutions of the Verhulst equation at $n \geq 0$.

Let us pose the Cauchy problem for the logistic equation. The selection of one or another branch is determined by the choice of the initial condition. We will set the initial condition $n = n_0$ at $t=0$ and we will look for a solution at $t>0$. It is easy to make sure that for $n_0 < n_\infty$ ($n_\infty < n_0$) the solution to the problem will be the red (green) branch in Figure 1. Thus, in the physics of aftershocks, when setting the Cauchy problem, one should set the initial conditions under the additional constraint $n_0 > n_\infty$. Moreover, it is reasonable to use the strong inequality

$$n_0 >> n_\infty = \gamma/\sigma. \tag{12}$$

Indeed, for $t \to \infty$, the frequency of aftershocks asymptotically approaches from above to the background (equilibrium) value $n_\infty$. Experience shows that as a rule $n_0 >> n_\infty$ after a strong earthquake.

An analysis of equation (10) under the condition $n_0 >> n_\infty$ indicates that at the first stages of evolution, the frequency of aftershocks decreases with time in accordance with the Omori formula (1). It is natural to call this stage of evolution "the Omori epoch". It is easy to see that the duration of the Omori epoch is limited by the inequality

$$t << 1/\sigma n_\infty. \tag{13}$$

The Omori epoch, the existence of which is predicted by our theory, was actually observed in many cases experimentally [18, 19].

### 4. Propagation of aftershocks

We come to the most difficult part of our search. Our challenge is, to express the propagation rate of aftershocks through the parameters of the master equation. Recall that the propagation phenomenon was discovered in [1] experimentally by analyzing the ribbed structure of aftershocks in the *x-t* plane (see also [2, 20]).

The nonlinear diffusion equation is interesting in many ways, but one of its properties is extremely important in the context of our problem. This property was discovered in [10, 11] and consists in the fact that the equation has self-similar solutions in the form of a traveling wave

$$n(x,t) = n(x \pm Ut). \tag{14}$$

It is this circumstance that played a decisive role in our choice of the KPP equation as the master equation. The estimation of the wave propagation velocity can be done by analyzing the dimensions of the coefficients of the master equation:

$$U \sim \sqrt{\gamma D}. \tag{15}$$

A set of particular solutions of the one-dimensional nonlinear diffusion equation in the form of a traveling wave is given in the book [21]. As an example, let us point out one of the particular solutions of this kind:

$$n(x,t) = n_\infty \left\{ -1 + C \exp\left[ \sqrt{\frac{\gamma}{6D}} (\pm x - Ut) \right] \right\}^{-2}. \tag{16}$$

Here $C$ is an arbitrary constant, $U = 5\sqrt{D\gamma/6}$.

So, we have outlined a promising direction for the theoretical study of the propagation of aftershocks, which was discovered earlier in the experiment. In this regard, it should be emphasized that the possibilities of analytical research are rather limited here. The most suitable are numerical methods for solving the nonlinear diffusion equation. In this case, known exact solutions of the type (16) can be effectively used to improve numerical schemes and to test the result of computational procedures. Numerical experiments will make it possible to study the propagation of aftershocks under various boundary and initial conditions and for various combinations of the problem parameters $\gamma$, $\sigma$, and $D$. This will certainly improve our understanding of the evolution of aftershocks in time and space.

## 5. Discussion

Valerio Faraoni [22] rewrote equation (2) in the following form:

$$\left(\frac{\dot{n}}{n}\right)^2 = \sigma^2 n^2. \tag{17}$$

This allowed him to see an interesting analogy between the evolution of aftershocks and the evolution of the Universe within the framework of the Friedman cosmological model.

Equation (3), rewritten as

$$\int_0^t \sigma(t')dt' = g(t), \tag{18}$$

can be used to formulate the inverse problem of an earthquake source. Here $g(t) = [n_0 n(t)]^{-1}[n_0 - n(t)]$ is a function known from the experiment. The essence of the inverse problem is to determine the deactivation coefficient $\sigma(t)$ as a function of time for a given function $g(t)$. The solution of the inverse problem for several dozen events revealed the existence of Omori epochs, during which $\sigma = \text{const}$ [18, 19].

After the end of the Omori epoch, the deactivation coefficient undergoes complex variations. Among the causes for the deviation from the Omori law, it is necessary to highlight the impact of endogenous and exogenous triggers on the earthquake source [23]. Formally, the effect of triggers can be taken into account in the framework of the relaxation theory of deactivation [6]. The deactivation equation has the form

$$\frac{d\sigma}{dt} = \frac{\bar{\sigma}(t) - \sigma}{\tau} + \xi(t), \qquad (19)$$

which is similar to that used to describe the mean temperature of the Earth's surface [24]. Here, $\tau$ is the characteristic time takes to approach the quasi-equilibrium state $\bar{\sigma}(t)$. The function $\xi(t)$ simulates the triggers. This function can be impulsive (round-the-world seismic echo [23, 25. 26], solar flares [27, 28]), sinusoidal (free oscillations of the Earth [29, 30], geomagnetic storms [31, 32]), or stochastic (seismic noise in the earthquake source).

We found that when setting the Cauchy problem for aftershocks, the following constraint should be imposed on the initial condition: $n_0 > n_\infty = \gamma / \sigma$. In this connection, the idea arises that the opposite inequality may be of interest in the phenomenological description of foreshocks. The issue requires additional study, since it is not entirely clear how to simulate the occurrence of the main shock in the process of increasing foreshock intensity along the logistic curve. Recall, however, that there are so-called earthquake swarms. The swarm is missing the main shock. Suppose that, for one reason or another, the parameter $n_\infty$ rises abruptly, and after a while returns to its original value. This kind of switching on/off will lead to parametric modulation of the earthquake frequency, and after switching on, the growth of $n(t)$ will occur along the lower branch, and after switching off, the decline of $n(t)$ will occur along the upper branch of the logistic function (see Figure 1). This is a preliminary scenario for the occurrence of an earthquake swarm.

Let us recall that the purpose of this work is to analyze the possibilities of a phenomenological description of earthquakes. We have chosen the KPP equation as the basis for this description. The question naturally arises whether it is possible within the framework of our ideas to make a prediction that could be tested experimentally? Within the framework of the model of earthquake swarms described above, there is the following quite definite prediction: the envelope of the frequency $n(t)$ of tremors in the swarm should have a left-side asymmetry relative to the top of the envelope. Further, the KPP equation has solutions in the form of a traveling wave at $n < n_\infty$ and $n > n_\infty$ [21]. This suggests that the phenomenon of propagation of epicenters with the velocity $U$ may be observed in the earthquake swarm.

Now we would like to touch upon the difficult question of the possibility of describing the main shock using the nonlinear diffusion equation (4). Figure 2 shows a fragment of the so-called Atlas of Aftershocks [18, 19]. We see that in three of the eight events shown in the figure, the deactivation factor becomes negative for a while. Let us introduce the notation

$$<|\sigma|> = \frac{1}{t_2 - t_1} \int_{t_1}^{t_2} |\sigma(t')| dt'. \qquad (20)$$

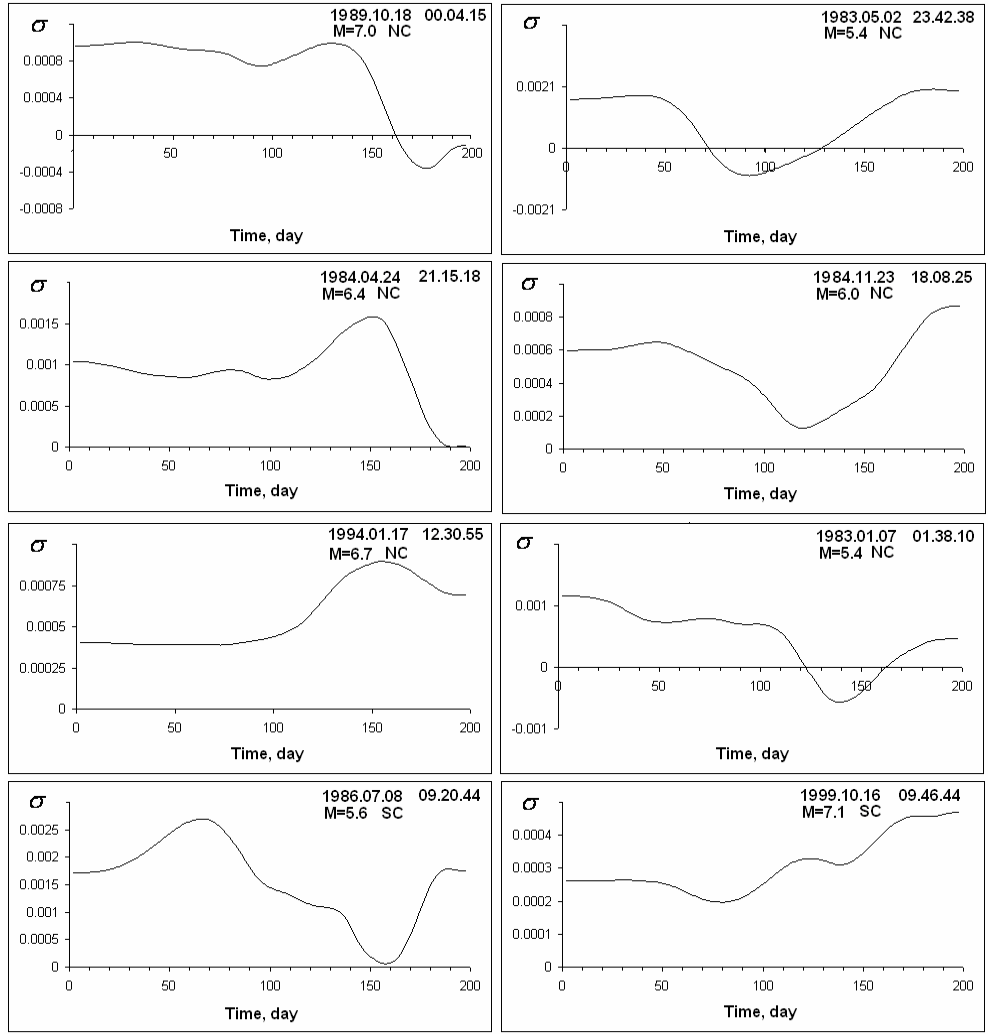

Fig. 2. Fragment of the Atls of Aftershocks, which gives an idea of the evolution of deactivation coefficient of the earthquake source.

For $\sigma < 0$ instead of (3) we will have

$$n(t) = \frac{n_0}{1 - n_0 <|\sigma|> (t - t_1)}. \qquad (21)$$

We see that at $t_2 \geq t_* = t_1 + \triangle t_*$ a singularity arises, which can be tried to represent as an image of the main shock. The value of $\triangle t_* = (n_0 <|\sigma|>)^{-1}$ is naturally called the waiting time for the main shock. If $t_2 < t_*$, then the main impact does not occur despite the fact that for some time the deactivation coefficient becomes negative.

To eliminate the singularity at $\sigma < 0$, the third term should be retained when expanding the function $F(n)$ in a power series: $F(n) = \gamma n + |\sigma|n^2 - \beta n^3$. It is assumed here that $\beta > 0$. Then $n_{nax} \approx |\sigma|/\beta$ for $t = t_*$, if for simplicity we put $\beta << |\sigma|^2/\gamma$. In the theory of dynamical systems, the described situation is called explosive instability. Thus, the theory predicts the occurrence of a geotectonic explosion in the form of the main shock of an earthquake when the sign of $\sigma$ changes. However, the reason for the possible change in sign is not yet clear to us.

An important note concerns the choice of unknown function in the equation (4). If we try to develop the phenomenological theory of the main shock within the framework of the Cauchy problem for master equation according to the scheme we outlined above, then it is reasonable to choose the energy, and not the frequency of earthquakes as the sought function.

And the last remark brings us back to the paper by Valerio Faraoni [22]. It presents a model of the Big Bang that led to the formation of the visible part of the Universe. Let's pay attention to the following. Faraoni's scenario distantly resembles the way we imagine the formations of the main shock, leading to the formation of aftershocks flow.

## 6. Conclusion

For the phenomenological description of aftershocks, we have proposed mathematical models, to one degree or another related to the nonlinear diffusion equation of Kolmogorov-Petrovsky-Piskunov. The choice of KPP as the governing equation is the most radical provision of the work presented here. Special attention was paid to the Omori law and the phenomenon of the propagation of aftershocks. A number of related issues were raised, including the question of the formation of earthquake swarms.

The authors are fully aware that the results of the work are preliminary, and that only the future will show how productive it is to represent the evolution of aftershocks on the basis of the nonlinear diffusion equation.

***Acknowledgments***. We express our deep gratitude to A.D. Zavyalov and O.D. Zotov for numerous fruitful discussions. We sincerely thank A.S. Potapov for his interest in our work. The work was supported by the project of the RFBR 18-05-00096, as well as the state assignment program of the IPhE RAS.

**References**

1. *Zotov O.D., Zavyalov A.D., Klain B.I.* On the spatial-temporal structure of aftershock sequences // In: Yanovskaya T. et all. (eds). Problems of Geocosmos–2018. Springer Proceedings in Earth and Environmental Sciences. 2020. Springer, Cham, P. 199–206.
2. *Zavyalov A.D., Guglielmi A.V., Zotov O.D.* Three problems of aftershock physics // J. Volcanology and Seismology. 2020. Vol. 14. No. 5. P. 341–352 // arXiv:2001.10569 [physics.geo-ph].
3. *Davison Ch.* The founders of seismology // Cambridge: University Press. 1930.
4. *Guglielmi A.V.* Omori's law: a note on the history of geophysics // Phys. Usp. 2017. V. 60. P. 319–324. DOI: 10.3367/UFNe.2017.01.038039.
5. *Omori F.*, On the aftershocks of earthquake // J. Coll. Sci. Imp. Univ. Tokyo, 1894, V. 7, P. 111–200.
6. *Guglielmi A.V., Zavyalov A.D.* The 150th anniversary of Fusakichi Omori // arXiv:1803.08555 [physics.geo-ph] // J. Volcanology and Seismology. 2018. V. 12. No. 5. P. 353–358 // IASPEI Newsletter, September 2018. P. 5–6.
7. *Hirano R.* Investigation of aftershocks of the great Kanto earthquake at Kumagaya // Kishoshushi. Ser. 2. 1924. V. 2. P. 77–83.
8. *Utsu T., Ogata Y., Matsu'ura R.S.* The centenary of the Omori formula for a decay law of aftershock activity // J. Phys. Earth. 1995. V. 43. P. 1–33.
9. *Guglielmi A.V.* Interpretation of the Omori law // arXiv:1604.07017 [physics.geo-ph] // Izv., Phys. Solid Earth. 2016. V. 52. No. 5. P. 785–786. doi:10.1134/S1069351316050165.
10. *Kolmogorov, A.N., Petrovsky, I.G., Piskunov, N.S.* Investigation of the equation of diffusion combined with increasing of the substance and its application to a biology problem // Bulletin of Moscow State University Series A: Mathematics and Mechanics. 1937. No. 1. P. 1–26.
11. *Fisher R.A.* The wave of advance of advantageous genes // Annual Eugenics. 1937. V. 7. P. 355–369.
12. *Ataullakhanov F.I., Zarnitsyna V.I., Kondratovich A.Yu., Sarbash V.I.* A new class of stopping self-sustained waves: a factor determining the spatial dynamics of blood coagulation // Phys. Usp. 2002. V. 45, P. 619–636.
13. *Murray J. D.* Mathematical biology. V.1. An Introduction // Springer. 2002. 551p.)
14. *Murray J. D.* Mathematical Biology. V.2. Spatial Model and Biomedical Applications // Springer. 2003. 839 p.

15. *Zeldovich Ya.B., Ruzmaikin A.A., Sokoloff D.D.* Magnetic Field in Astrophysics. New York: Gordon and Breach. 1983.
16. *Kuzmin Y.O.* Recent geodynamics and slow deformation waves // Izv., Phys. Solid Earth. 2020. V. 56. P. 595–603.
17. *Verhulst, P.F.* Notice sur la loi que la population poursuit dans son accroissement. Correspondance mathématique et physique. 1838. V. 10. P. 113—121.
18. *Guglielmi A.V., Zotov O.D., Zavyalov A.D.* A project for an Atlas of aftershocks following large earthquakes // J. Volcanology and Seismology. 2019. V. 13. No. 6. P. 415–419. DOI: 10.1134/S0742046319060034.
19. *Guglielmi A.V., Zotov O.D., Zavyalov A.D.* Atlas of aftershock sequences of strong earthquakes // In: Yanovskaya T. et all. (eds). Problems of Geocosmos–2018. Springer Proceedings in Earth and Environmental Sciences. 2020. Springer, Cham. P. 193–198.
20. *Guglielmi A.V., Zotov O.D.* Aftershocks of the 2013 deep Okhotsk earthquake // arXiv:2007.14754 [physics.geo-ph].
21. *Polyanin A.D., Zaitsev V.F.* Handbook of nonlinear partial differential equations, 2nd Edition // Chapman & Hall/CRC Press, Boca Raton. 2012.
22. *Faraoni V.* Lagrangian formulation of Omori's law and analogy with the cosmic Big Rip // Eur. Phys. J. C 2020. 80:445 https://doi.org/10.1140/epjc/s10052-020-8019-2
23. *Guglielmi A.V.* Foreshocks and aftershocks of strong earthquakes in the light of catastrophe theory // Physics-Uspekhi. 2015, V. 58. No. 4. P. 384–397. DOI: https://doi.org/10.3367/UFNr.0185.201504f.0415 .
24. *Byalko A.V.* Relaxation theory of climate // Physics–Uspekhi. 2012. V. 55. No. 1. P. 103–108.
25. *Guglielmi A.V., Zotov O.D., Zavyalov A.D.* The aftershock dynamics of the Sumatra–Andaman earthquake // Izv. Phys. Solid Earth. 2014. V. 50. No. 1. P. 64–72.
26. *Zotov O.D., Zavyalovb A.D., Guglielmi A.V., Lavrov I.P.* On the possible effect of round-the-world surface seismic waves in the dynamics of repeated shocks after strong earthquakes // Izv. Phys. Solid Earth. 2018. V. 54. No. 1,. P. 178–191.
27. *Tarasov N.T.* The effect of solar activity on the seismicity of the Earth // Inzhenernaya fizika [Engineering Physics]. 2019. No. 6. P. 23–33. (In Russian)
28. *Guglielmi A.V., Klain B.I.* Effect of the Sun on Earth's seismicity // Solar-Terrestrial Physics. 2020. V. 6, No. 1. P. 89–92. DOI: 10.12737/stp-61202010 // arxiv.org/abs/1909.00879.
29. *Guglielmi A., Zotov O.* Impact of the Earth's oscillations on the earthquakes // arXiv:1207.0365v1 [physics.geo-ph].
30. *Guglielmi A.V., Zotov O.D.* On the near-hourly hidden periodicity of earthquakes // Izv. Phys. Solid Earth. 2013. V. 49. No. 1. P. 1–8. DOI: 10.1134/S1069351313010047.
31. *Buchachenko A.L.* Microwave stimulation of dislocations and the magnetic control of the earthquake core // Physics-Uspekhi. 2019, V. 62, No. 1, P. 46–53. DOI: 10.3367/UFNe.2018.03.038301.
32. *Guglielmi A.V. Klain B.I. Kurazhkovskaya N.A.* Earthquakes and geomagnetic disturbance // Solnechno-zemnaya fizika. 2020. Vol. 6. Iss. 4. P. 99–109. DOI: 10.12737/szf-54201911. (In Russian)